\documentclass[12pt]{article}
\usepackage{lmodern}
\usepackage{graphicx}
\usepackage{tikz}
\usetikzlibrary{arrows.meta,positioning,fit,backgrounds,calc}
\usepackage{amsmath,amssymb}
\usepackage{color}
\usepackage{url}
\usepackage{natbib}
\usepackage{fullpage}
\usepackage{authblk}
\usepackage[british]{babel}
\usepackage[utf8]{inputenc}
\usepackage{algorithm}
\usepackage{algorithmicx}
\usepackage{algpseudocode}
\usepackage[section]{placeins}
\usepackage{appendix}
\usepackage{orcidlink} 

\usepackage[a4paper, left=42pt, right=42pt, top=69pt, bottom=127pt]{geometry}

\newtheorem{proposition}{Proposition}
\newenvironment{proof}[1][Proof]{\par\noindent\textbf{#1.} }{\hfill$\square$\par\medskip}

\newcommand{\bfkappa}{\boldsymbol{\kappa}}
\newcommand{\bfzeta}{\boldsymbol{\zeta}}
\newcommand{\bftheta}{\boldsymbol{\vartheta}}
\newcommand{\bfxi}{\boldsymbol{\xi}}
\newcommand{\bfSigma}{\boldsymbol{\Sigma}}
\newcommand{\bfDelta}{\boldsymbol{\Delta}}
\def\bfa{\mathbf{a}}
\def\bfA{\mathbf{A}}

\def\bfd{\mathbf{d}}
\def\bfD{\mathbf{D}}

\def\bfs{\mathbf{s}}

\def\bfu{\mathbf{u}}
\def\bfU{\mathbf{U}}
\def\bfv{\mathbf{v}}
\def\bfV{\mathbf{V}}
\def\bfw{\mathbf{w}}

\def\bfX{\mathbf{X}}
\def\bfz{\mathbf{z}}
\newcommand{\bfzero}{\mathbf{0}}
\newcommand{\NN}{\mathcal{N}}
\newcommand{\RR}{\mathbb{R}}
\newcommand{\calA}{\mathcal{A}}
\newcommand{\calE}{\mathcal{E}}
\newcommand{\calG}{\mathcal{G}}
\newcommand{\calM}{\mathcal{M}}

\newcommand{\thc}{\bftheta_{\mathrm{c}}}
\newcommand{\thr}{\bftheta_{\mathrm{r}}}

\newcommand{\appfig}[2]{\IfFileExists{#1}{\includegraphics[#2]{#1}}{%
  \fbox{\parbox[c][5cm][c]{0.9\textwidth}{\centering\small file
  \texttt{\detokenize{#1}} not found}}}}
\providecommand{\keywords}[1]{\small\textbf{\textit{Keywords---}} #1}

\title{Latent space dynamic network actor models\\ with position-rate link}
\author{Alberto Caimo \orcidlink{0000-0001-8956-7166} $^{1}$}
\author{Isabella Gollini \orcidlink{0000-0002-7738-5688} $^{1,2}$}
\affil{\small $^1$School of Mathematics and Statistics; $^2$Rinn AI, University College Dublin, Ireland}
\date{}

\begin{document}
\maketitle

\abstract{Relational event data record time-stamped interactions between actors. Dynamic network actor models represent each interaction as two separate decisions: when an actor becomes active and whom they select. By modelling these decisions separately, they cannot establish whether the actors who interact most are also those most frequently selected by others.
To address this limitation, we propose a latent space dynamic network actor model in which sender and receiver positions capture receiver choice, while a position-rate link relates each actor's latent positions to their activity level. The link is based on two summaries that remain unchanged when the latent positions are updated: latent popularity, measuring how much an actor is favoured by others, and excess self-affinity, measuring the extent to which an actor favours similar actors beyond what is explained by their popularity. We adopt a Bayesian modular approach in which latent positions are estimated using receiver choices alone. We illustrate the model using e-mail data from a manufacturing company. The results show that employees who send the most e-mails are also those most frequently selected as recipients. When two or more latent dimensions are included, preference for colleagues with similar latent characteristics makes a smaller additional contribution. The latent positions improve receiver-choice prediction, while the position-rate link captures differences in actor activity.}

\keywords{Bayesian modular inference; dynamic network actor models; e-mail communication; latent space models; relational events}
\maketitle

\section{Introduction}
\label{sec:intro}

Relational event models describe the dynamics of directed interactions between actors as a function of past events \citep{but08,per:wol13,bia:fil:lom:wit24}. Dynamic network actor models (DyNAMs) \citep{sta:blo17} take an actor-oriented approach, in which each event results from two decisions: when an actor becomes active, modelled by a rate sub-process, and which other actor it selects, modelled by a choice sub-process. The two sub-processes have separate parameters. DyNAMs with actor-specific random effects \citep{uza:ama:sta23} account for heterogeneity in activity levels, but do not capture heterogeneity in actors' preferences for interaction partners. Latent space models \citep{hof:raf:han02,hof05,hof21} represent interaction preferences through unobserved actor positions. These approaches have also been extended to relational event data \citep{hua:sol:pau:xu22,art:wit23,ras:cor23}, but in these models the same latent quantities govern both how frequently and with whom actors interact. Within the DyNAM framework, latent positions can instead enter the choice sub-process, capturing actors' preferences for interaction partners separately from their activity levels, which are governed by the rate sub-process. Whether, and to what extent, these two aspects of actor behaviour are related is an important question. 

In the social relations model \citep{war:ken:sto79}, the covariance between the sender and receiver effects of an actor measures whether those who give more also receive more, and sociological accounts relate an actor's position in a network to how much it interacts, through cohesion \citep{col88} or brokerage \citep{bur92}. A DyNAM with separately
specified sub-processes has no parameter for such an association. The aim of this work is to connect the two sub-processes. 

The paper makes three main contributions. In Section~\ref{sec:model}, we presents latent space DyNAMs with a position-rate link: latent sender and receiver positions enter the choice sub-process, and summaries of each actor's positions enter the distribution of
its activity effect in the rate sub-process via an additive spline
regression. The choice model determines the positions only up to a group of
transformations that is larger than the rotations. We characterise this group
(Appendix~\ref{app:ident}) and build the link on summaries that it leaves
unchanged. 
Section~\ref{sec:inference} presents modular Bayesian inference and model
comparison: the positions are estimated from the choice of receivers alone,
and the link is estimated conditional on them
\citep{liu:bay:ber09,plu15,jac:mur:hol:rob17}, so that the summaries describe
preferences independently of activity. Section~\ref{sec:computation} describes
the computational approach and Section~\ref{sec:simulation} a simulation
study. In Section~\ref{sec:application} we analyse the e-mails exchanged among
the employees of a manufacturing company \citep{mic:pal:kaz11}: the employees
who write most are those whom the others prefer as recipients, an association
that DyNAMs with separate sub-processes cannot express, and a preference for
colleagues of one's own kind adds a smaller contribution once the latent space
has two dimensions. Section~\ref{sec:discussion} concludes with a discussion,
and the appendices provides identifiability results and robustness checks.

\section{The model}
\label{sec:model}

\subsection{A latent space DyNAM}
\label{sec:dynam}

Let $\calA=\{1,\dots,n\}$ be a set of actors observed over the interval
$[0,T]$, and let $\calE=\{(i_m,j_m,t_m)\}_{m=1}^{M}$ be a sequence of
directed events, with sender $i_m$, receiver $j_m\neq i_m$ and time $t_m$,
ordered in time. Each event is the outcome of two decisions: an actor becomes active, and when active it selects a receiver. The model describes these decisions using the following quantities. The statistics $\bfs(i;t)$ and $\bfs(i,j;t)$ are computed from the history of the process up to time $t$, for example, from the numbers of events sent and received so far: the first describes actor $i$, the second the pair $(i,j)$. The parameters $\bfxi$ and $\bfzeta$ are associated to the statistics $\bfs(i;t)$ and $\bfs(i,j;t)$, respectively. The activity effect $a_i$ is a random effect of actor $i$, so that $\exp(a_i)$ is the rate at which $i$ creates an event when its statistics are zero. Finally, $\bfu_i\in\RR^{D}$ and $\bfv_j\in\RR^{D}$ are the sender position of actor $i$ and the receiver position of actor $j$, in a latent space of dimension $D$.

Actor $i$ becomes active with intensity $\lambda_i(t)$, and when active it
selects receiver $j$ with probability $p_{ij}(t)$, where
\begin{equation*}
\lambda_i(t)=\exp\{\bfxi^\top\bfs(i;t)+a_i\},\qquad
p_{ij}(t)=\frac{\exp\{\bfzeta^\top\bfs(i,j;t)+\bfu_i^\top\bfv_j\}}
{\sum_{k\neq i}\exp\{\bfzeta^\top\bfs(i,k;t)+\bfu_i^\top\bfv_k\}}.
\end{equation*} 
The inner product includes a receiver effect
and a squared-distance specification as special cases \citep{hof21}.
With statistics that change only when events occur, the likelihood factorises into a rate part $L_{\mathrm r}(\bfxi,\bfa)$ and a
choice part $L_{\mathrm c}(\bfzeta,\bfU,\bfV)$:
$$\begin{aligned}
L(\bfxi,\bfa,\mathrm\bfzeta,\bfU,\bfV) &=L_{\mathrm r}(\bfxi,\bfa)\times L_{\mathrm c}(\bfzeta,\bfU,\bfV) \\
&= \prod_{m=1}^{M}\lambda_{i_m}(t_m)\exp\Big\{-\Delta t_m\sum_{i\in\calA}\lambda_i(t_m)\Big\}
\times\prod_{m=1}^{M}p_{i_mj_m}(t_m),
\end{aligned}$$
where $\Delta t_m=t_m-t_{m-1}$, with $t_0=0.$
The latent positions capture persistent preferences not accounted for by the observed statistics. Without them, persistent affinity between actors $i$ and $j$ may be attributed to inertia, reflecting the classical difficulty of distinguishing unobserved heterogeneity from state dependence \citep{hec81}. 
The positions also give summaries of each actor's preferences that are estimated from the choice of receivers alone. Observed measures, such
as the number of events an actor receives, partly reflect its own activity
through reciprocation, and relating them to activity would relate activity to
one of its consequences.

\subsection{Interpretation of the latent positions}
\label{sec:ident}

The choice probabilities depend on the positions only through the differences
$\bfu_i^\top(\bfv_j-\bfv_k)$ between alternative receivers. These do not change
if every $\bfu_i$ is replaced by $\bfA\bfu_i$ and every $\bfv_j$ by
$\bfA^{-\top}(\bfv_j+\bfw)$, for any invertible $D\times D$ matrix $\bfA$ and
any $\bfw\in\RR^D$, and the data determine the positions up to these
transformations and no further (Appendix~\ref{app:ident}). 

The prior on the latent positions is invariant only to transformations such as rotations and reflections, leaving other features, such as vector lengths and receiver locations, determined by the model specification rather than by the data. We therefore build the link only on summaries that these transformations leave unchanged. Let $\bar\bfu$ and $\bar\bfv$ be the average sender and receiver positions and
$\bfd_i=\bfv_i-\bar\bfv$. The \emph{self-affinity} of actor $i$,
$r_i=\bfu_i^\top\bfd_i$, is the latent log-odds that $i$ selects an actor at
its own receiver position rather than one at the centroid. A positive $r_i$
means that $i$ tends to select actors of its own kind, a negative $r_i$ actors
unlike itself. It combines two properties: an actor may favour its own kind
because every sender favours actors like it, or because of a preference of its
own. Accordingly,
\begin{equation*}
r_i=\pi_i+\rho_i,\qquad
\pi_i=\bar\bfu^\top\bfd_i,\qquad
\rho_i=(\bfu_i-\bar\bfu)^\top\bfd_i,
\end{equation*}
where $\pi_i$ is the \emph{latent popularity} of $i$, how much the average
sender favours actors at its position, and $\rho_i$ is its \emph{excess
self-affinity}, how much its own preference departs from that of the average
sender. Both terms are invariant.
The latent popularity $\pi_i$ is the log-odds that an average sender selects $i$ rather than an average receiver: it is a feature of $i$ as a receiver, the latent counterpart of a receiver effect. The excess self-affinity $\rho_i$ is positive when $i$ prefers its own kind more than the average sender does, and negative when it selects it less. An association between activity and $\pi_i$ concerns how an actor is seen by the others, and corresponds to the covariance between sender and receiver effects in the social relations model \citep{war:ken:sto79,hof21}; an association with $\rho_i$ concerns preferences that differ between senders, which only latent positions express.

With one latent dimension the positions are scalars, determined up to a
common change of sign. In a plot of $u_i$ against $v_i-\bar v$, actors in the
first and third quadrants have positive self-affinity, and latent popularity
is proportional to the centred receiver position. For actors whose $u_i$ has
the sign of $\bar u$, self-affinity is popularity rescaled by the strength of
their preferences, $r_i=(u_i/\bar u)\,\pi_i$. With more dimensions the
senders can favour different receivers.

\subsection{The position-rate link}
\label{sec:linkspec}

\begin{figure}[t]
\centering
\appfig{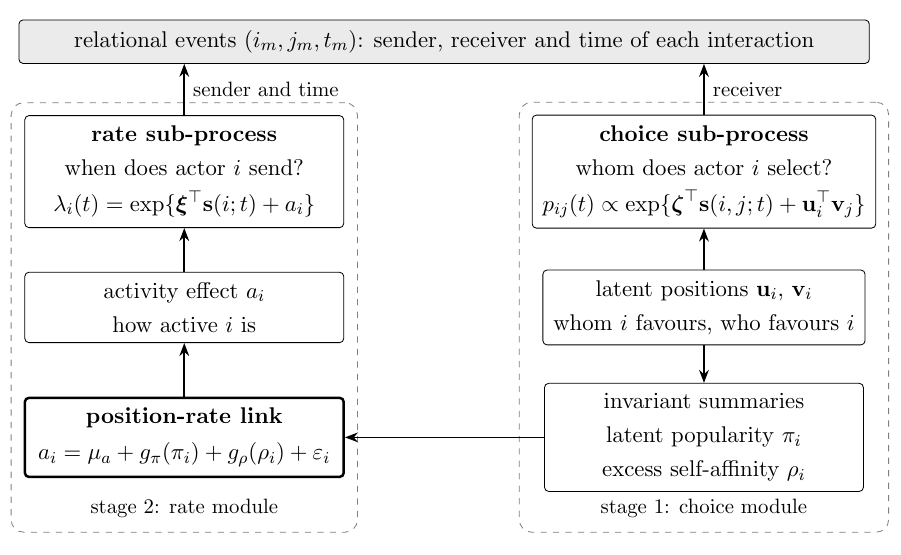}{width=0.9\textwidth}
\caption{Structure of the latent space DyNAM with position-rate link. The
choice sub-process depends on the latent positions, the rate sub-process on
the activity effects, and the link relates the activity effects to invariant
summaries of the positions. The dashed boxes are the two stages of modular
inference (Section~\ref{sec:modular}).}
\label{fig:diagram}
\end{figure}

Let $\tilde\pi_i=\pi_i - \bar\pi$ and $\rho_i-\bar\rho$ be the centred position summaries, we define the position-rate link as a second-level regression
\begin{equation*}
a_i=\mu_a+g_\pi(\tilde\pi_i)+g_\rho(\tilde\rho_i)+\varepsilon_i,\qquad
\varepsilon_i\sim\NN(0,\sigma_a^2),
\end{equation*}
in which each component $g$ is a natural cubic spline with $K$ basis
functions and knots at quantiles of the summary, whose parameters 
$\bfkappa$ follow Normal distributions. The component $g_\pi$ describes
how the expected activity effect varies with popularity when excess
self-affinity is held fixed: $\exp\{g_\pi(x)-g_\pi(x')\}$ is the ratio of the
typical rates of two actors with the same statistics that differ only in
popularity. We summarise each component by its \emph{contrast}, the
difference between its values at the 90\% and 10\% quantiles of the summary \citep{har15}.
The parameter $\sigma_a$ measures the variation in activity that the link
does not explain, and $\bfkappa=\bfzero$ gives the usual exchangeable activity
effects. The components are not restricted to be linear or monotone, since
the arguments that relate position to activity do not agree on a direction.
A link in self-affinity alone, $g(\tilde r_i)$, would force the two
components to act through their sum. Figure~\ref{fig:diagram} summarises the
model.

\subsection{Interpretation of the link} The link replaces the common prior mean of
the activity effects by an actor-specific mean
$m_i=\mu_a+g_\pi(\tilde\pi_i)+g_\rho(\tilde\rho_i)$. The events of actor $i$
alone give the estimate $\hat a_i=\log(N_i/E_i)$, where $E_i=\sum_m\Delta t_m\exp\{\bfxi^\top\bfs(i;t_m)\}$ is its exposure, with precision approximately $N_i$ which is the number of events sent by actor $i$. Combining the two shows
(Section~\ref{sec:computation}) that the posterior mode of $a_i$, given the
other unknowns, is approximately
\begin{equation}
\label{eq:shrink}
\tilde a_i\approx w_i\,m_i+(1-w_i)\,\hat a_i,\qquad
w_i=\frac{1}{1+\sigma_a^2N_i}.
\end{equation}
The activity effect of an actor who has sent many events is therefore
determined by its own events, and the link changes only the effects of actors
with few events, which are shrunk towards the link rather than towards a
common mean. What the link adds is the function $g$ and the residual
variation $\sigma_a$: an estimate, with its uncertainty, of how activity
varies with position. Unlike a regression of estimated activity effects on
estimated summaries after the fact, it takes into account that the precision
of the activity effects varies by orders of magnitude between actors. The
link describes an association between characteristics of the same actors; it
does not imply that positions affect activity.

\section{Inference}
\label{sec:inference}

\subsection{Modular inference}
\label{sec:modular}

Let $\thc=(\bfzeta,\bfU,\bfV)$ contain the parameters and latent positions of the choice module and
$\thr=(\bfxi,\bfa,\mu_a,\bfkappa)$ the parameters of the rate module. With the link,
the positions enter both modules, so that in the joint posterior the event
times would also inform the positions: configurations whose summaries predict
activity well would receive higher posterior probability. We want the
summaries to describe the choice of receivers alone, so that the link relates
two separately defined quantities and a misspecified link cannot distort the
positions \citep{liu:bay:ber09}. We therefore use the cut posterior
\citep{plu15,jac:mur:hol:rob17}
\begin{equation*}
p_{\mathrm{cut}}(\thc,\thr\mid\calE)=p_{\mathrm c}(\thc\mid\calE)\times
p_{\mathrm r}\left(\thr\mid\calE,\bfX(\thc)\right),
\end{equation*}
where $p_{\mathrm c}$ is the posterior of the choice module alone and
$p_{\mathrm r}$ that of the rate module when the centred summaries
$\bfX(\thc)$ are treated as covariates. 

Inference proceeds in two stages (Figure~\ref{fig:diagram}): the choice module is estimated first, and the rate module is then estimated given the summaries of the estimated positions. Our main estimate is the so-called ``plug-in'' posterior, which fixes the summaries at their values evaluated at the posterior mean positions. The ``cut'' posterior also propagates the uncertainty about the positions, but this adds a measurement error that attenuates the components; we use it to check that the features of the link do not result from ignoring that uncertainty.

We set zero-centred Normal prior distributions with variance 4 for $\bfxi$ and $\bfzeta$, 100 for $\mu_a$, 1 for each coordinate of the positions, and 1 for the parameters of the link, whose basis functions are scaled to unit standard deviation across actors. The residual standard deviation $\sigma_a$ is set by maximising the marginal likelihood of the rate module \citep{wood:fas:sta11}. Each module
is simulated by a Metropolis-adjusted Langevin algorithm (MALA) \citep{rob:twe96}
preconditioned with the curvature at the posterior mode; since the posterior
of the choice module can have several local modes, its mode is searched from multiple starting configurations (Section~\ref{sec:computation}).

\subsection{Model comparison}
\label{sec:assessment}

We compare three nested models: $\calM_0$, the DyNAM with activity effects and no latent positions; $\calM_1$, the latent space DyNAM without link; and $\calM_2$, the DyNAM with the link. $\calM_1$ and $\calM_0$ differ in the
choice sub-process, whereas $\calM_2$ and $\calM_1$ differ only in the
distribution of the activity effects. The evidence for the link is measured
by the Bayes factor between $\calM_2$ and $\calM_1$, computed from Laplace
approximations to the marginal likelihoods of the rate module
\citep{tie:kad86}, with $\sigma_a$ set in each model. The same comparison
between a link with both components and a link with one of them gives the
evidence for one component given the other, for example for excess
self-affinity given popularity. We also report the proportion $R^2_a$ of the
variation of the activity effects between actors that the link explains
\citep{gel:par06}. Predictive performance is measured on held-out events by
the logarithmic score \citep{gne:raf07}, which separates into a choice score,
for the receivers, and a rate score, for the sending times.

\section{Computational approach}
\label{sec:computation}

\subsection{Choice module} Since the log-posterior is not concave in $(\bfU,\bfV)$, so we maximise it by using the BFGS algorithm with analytic gradients from several random starting points and one spectral configuration (from the singular value decomposition of the row-centred matrix of $\log(1+\text{events from } i \text{ to } j)$), and keep the best mode. Let $\bfSigma_{\mathrm c}$ be the inverse of the Hessian of the negative log-posterior at the mode, where eigenvalues below half the prior precision of the positions are raised to that value. This gives a large but finite variance to the directions along rotations and translations, in which the posterior is flat or determined by the prior.
MALA proposes
$\thc'=\thc+\tfrac{\epsilon^2}{2}\bfSigma_{\mathrm c}\nabla\log
p_{\mathrm c}(\thc\mid\calE)+\epsilon\,\bfSigma_{\mathrm c}^{1/2}\bfz$, with
$\bfz$ following a standard Normal, so that steps are long where the posterior is wide and short where it is narrow. The step size $\epsilon$ is tuned during burn-in to an acceptance rate near 0.57 \citep{rob:ros98}. Since the posterior is unchanged by rotations, each draw is aligned with the mode by the orthogonal Procrustes rotation (a change of sign when $D=1$), and for $D\ge2$ the pre-conditioner is rotated with each configuration.

\subsection{Rate module} The rate statistics of an actor change only when it
sends or receives an event, so that the likelihood and its derivatives cost
$O(M+n)$ computations. Conditional on the summaries and $\sigma_a$, the log-posterior for $(\bfxi,\bfa,\mu_a,\bfkappa)$ is strictly concave. We locate its mode using Newton's method and jointly sample the parameters using MALA, preconditioned by the inverse Hessian. The parameter $\sigma_a$ maximises the Laplace approximation to the marginal likelihood of the rate module. The spline knots are computed once from the plug-in summaries.

\subsection{Cut posterior} For each of the $S=300$ thinned draws of the choice
module, the summaries are recomputed, the bases are evaluated at the fixed
knots, and one draw is taken from the Laplace approximation to the rate
module given these summaries, which is adequate because this posterior is
log-concave and close to Normal. Algorithm~\ref{alg:twostage} summarises the estimation procedure.

\begin{algorithm}[h]
\caption{Two-stage inference}
\label{alg:twostage}
\begin{algorithmic}[1]
\State Choice module: 
\Statex (i) Find posterior mode by BFGS from multiple starts; \Statex (ii) Compute $\bfSigma_{\mathrm c}$; 
\Statex (iii) Sample by MALA, aligning each draw with mode
\Statex (iv) Compute the plug-in summaries from posterior means of aligned positions 
\Statex (v) Centre positions and build the spline bases
\State Rate module: 
\Statex (i) Set $\sigma_a$ by maximum marginal likelihood; 
\Statex (ii) Find mode and sample plug-in posterior by MALA
\Statex (iii) Cut posterior: for draws of choice module, recompute summaries 
\Statex (iv) Draw from Laplace approximation to rate module
\end{algorithmic}
\end{algorithm}

\subsection{Shrinkage of the activity effects} Given $\mu_a$, $\bfkappa$, $\sigma_a$ and $\bfxi$, the link gives $a_i$ the prior distribution $\NN(m_i,\sigma_a^2)$, and the events of actor $i$ contribute the log-likelihood $N_ia_i-E_i\exp(a_i)$. For $N_i>0$ this is maximised at
$\hat a_i=\log(N_i/E_i)$, where its second derivative is $-N_i$, so that its
second-order Taylor series expansion around $\hat a_i$ is
$-\tfrac12N_i(a_i-\hat a_i)^2$ up to a constant. Combined with the prior, it
gives an approximately Normal conditional posterior distribution with mode
$(m_i/\sigma_a^2+N_i\hat a_i)/(1/\sigma_a^2+N_i)$, which is \eqref{eq:shrink}.

\section{Simulation study}
\label{sec:simulation}

We simulated relational events from the model for $n=60$ actors, with the statistics of
the application and $\xi=0.1$, $\zeta=0.5$. The sender positions were drawn
from the Normal distribution with mean $(1,0,\dots,0)^\top$ and identity
covariance matrix, so that the senders share a
direction of preference, and the receiver positions from the standard Normal
distribution. The activity effects were
$a_i=g_\pi(\pi_i)+g_\rho(\rho_i)+\varepsilon_i$, with
$\varepsilon_i\sim\NN(0,0.5^2)$, where each component is either zero or the
smooth step $1.5\{\mathrm{logit}^{-1}(3x)-1/2\}$ of the centred summary. 

Each replicate draws new positions and activity effects and simulates events.
The design is as follows.
\begin{itemize}
\item 4 scenarios: activity is associated with neither summary, with
popularity only, with excess self-affinity only, or with both.
\item Latent space dimensions: $D=1$ or $D=2$.
\item Number of events: $M=2{,}000$ or $8{,}000$, about 33 or 133 interactions sent
per actor.
\item Replicates: 20 for each combination of scenario, $D$ and $M$.
\item Rate statistic: $s(i;t) = \log\{1+N^{\mathrm{in}}_i(t)\}$, where
$N^{\mathrm{in}}_i(t)$ is the number of e-mails received by $i$ up to time $t.$
\item Choice statistic: $s(i,j;t) = \log\{1+N_{ji}(t)\}$, where
$N_{ji}(t)$ is the number of e-mails sent by $j$ to $i$ up to time $t.$
\item Link spline: $K=4$ basis functions for each component.
\end{itemize}
The specification is deliberately minimal, with one statistic in each
sub-process, so that the positions describe the stable structure of the
choices beyond reciprocation.

The estimated summaries are close to the true ones. With $M=2{,}000$ the correlation between estimated and true values is between 0.95 and 0.97 for latent popularity and between 0.86 and 0.93 for excess self-affinity. With $M=8{,}000$ it is at least 0.98 for latent popularity and at least 0.96 for excess self-affinity.

With $M=8{,}000$ the contrasts of both components are recovered with little bias (Table~\ref{tab:sim}). With $M=2{,}000$ they are attenuated, that is, shrunk towards zero: under the plug-in posterior by 2\% to 6\% for popularity and by 13\% for excess self-affinity, and by more under the cut posterior. The 95\% credible intervals for the contrasts contain the true value in at least 90\% of the replicates. The exception is excess self-affinity with $M=2{,}000$ when both components are present, where the coverage is 75\% to 80\%.

The Bayes factors do not signal association that is not present. When activity is associated with neither summary, the Bayes factor for the link is never greater than 3. When only popularity is present, the Bayes factor for excess self-affinity given popularity never exceeds 3 either, so the shrinkage of the sender positions does not create spurious evidence for excess self-affinity. A component that is present is detected in most replicates, except excess self-affinity with $M=2{,}000$ when popularity is present too.

\begin{table}[ht]
\centering
\caption{Simulation study (20 replicates per row). Contrast: true value and
mean over the replicates of the posterior mean under the plug-in posterior.
Bayes factor (BF) $>3$: proportion of the replicates in which the Bayes factor
for popularity given excess self-affinity ($\pi\mid\rho$), or for excess
self-affinity given popularity ($\rho\mid\pi$), exceeds 3.}
\label{tab:sim}
\smallskip
{
\begin{tabular}{@{}lccrrrrrr@{}}
\hline
 & & & \multicolumn{2}{c}{Contrast, $\pi$} & \multicolumn{2}{c}{Contrast, $\rho$} & \multicolumn{2}{c}{BF $>3$}\\
Association & $D$ & $M$ & true & estimate & true & estimate & $\pi\mid\rho$ & $\rho\mid\pi$\\
\hline
None & 1 & 2,000 & 0.00 & 0.03 & 0.00 & 0.06 & 0.00 & 0.00\\
 & 1 & 8,000 & 0.00 & 0.00 & 0.00 & 0.06 & 0.00 & 0.00\\
 & 2 & 2,000 & 0.00 & 0.02 & 0.00 & $-0.06$ & 0.00 & 0.00\\
 & 2 & 8,000 & 0.00 & 0.07 & 0.00 & 0.08 & 0.00 & 0.00\\[5pt]
$\pi$ only & 1 & 2,000 & 1.41 & 1.29 & 0.00 & 0.00 & 0.95 & 0.00\\
 & 1 & 8,000 & 1.41 & 1.45 & 0.00 & 0.02 & 1.00 & 0.00\\
 & 2 & 2,000 & 1.40 & 1.38 & 0.00 & $-0.03$ & 0.95 & 0.00\\
 & 2 & 8,000 & 1.41 & 1.42 & 0.00 & 0.03 & 1.00 & 0.00\\[5pt]
$\rho$ only & 1 & 2,000 & 0.00 & 0.09 & 1.30 & 1.21 & 0.00 & 0.80\\
 & 1 & 8,000 & 0.00 & 0.07 & 1.32 & 1.28 & 0.00 & 0.95\\
 & 2 & 2,000 & 0.00 & $-0.01$ & 1.47 & 1.37 & 0.00 & 0.95\\
 & 2 & 8,000 & 0.00 & 0.03 & 1.45 & 1.48 & 0.00 & 1.00\\[5pt]
$\pi$ and $\rho$ & 1 & 2,000 & 1.41 & 1.35 & 1.36 & 1.10 & 1.00 & 0.55\\
 & 1 & 8,000 & 1.41 & 1.46 & 1.34 & 1.33 & 1.00 & 0.95\\
 & 2 & 2,000 & 1.40 & 1.37 & 1.46 & 1.18 & 0.90 & 0.55\\
 & 2 & 8,000 & 1.40 & 1.48 & 1.45 & 1.44 & 1.00 & 1.00\\
\hline
\end{tabular}}
\end{table}

\section{Application to e-mail communication}
\label{sec:application}

\subsection{Data and specification}
\label{sec:app-data}

The data record the sender, the recipient and the time of 82,927 e-mails
exchanged among 167 employees of a mid-sized manufacturing company in Poland
between January and September 2010 \citep{mic:pal:kaz11}; the public data
contain no information on roles or departments. Each e-mail reflects the
sender's decisions about when to write and whom to contact, which matches the
actor-oriented structure of a DyNAM. We treat each sender-recipient pair as an
event, analyse the second fifth of the observation period and keep the employees 
who both send and receive within it. This leaves
$n=120$ employees and $M=16{,}256$ events over 53.6 days; the number of e-mails sent ranges from 2 to 818, with median 93.
Each sub-process is specified as in Section~\ref{sec:simulation}. We present the analysis with one latent dimension, in which the positions can be
displayed, and summarise the results with two and three dimensions in
Section~\ref{sec:app-D2}.

\subsection{Latent structure}
\label{sec:app-positions}

\begin{figure}[t]
\centering
\appfig{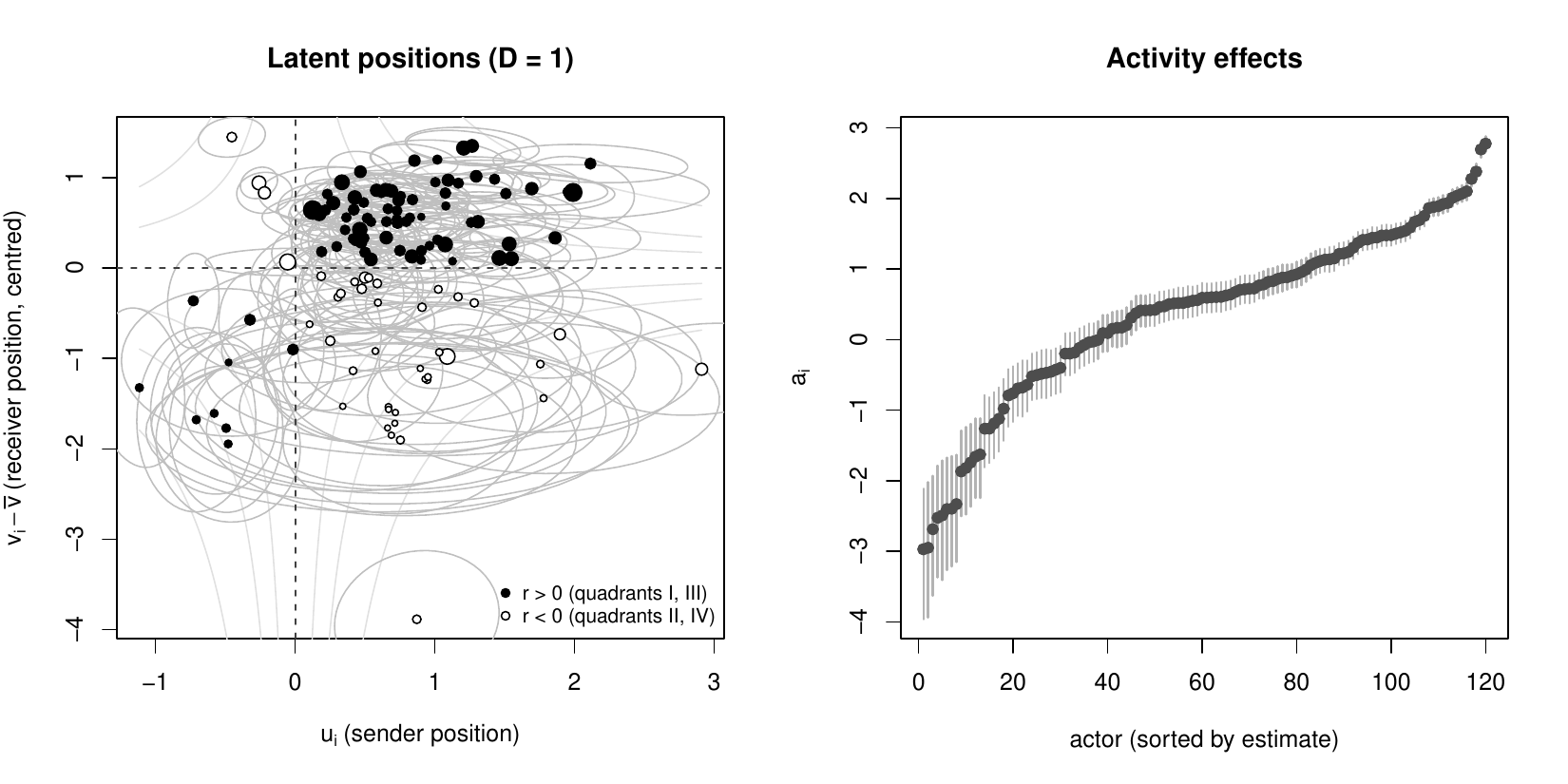}{width=\textwidth}
\caption{Left: posterior means of the sender position $u_i$ and of the centred receiver position $v_i-\bar v$, $D=1$, with 95\% posterior regions.
Filled points: positive self-affinity (first and third quadrants); open points: negative self-affinity. The size of a point is proportional to the number of e-mails sent. Right: posterior means and 95\% credible intervals of the activity effects (sorted).}
\label{fig:positions}
\end{figure}

The reciprocity parameter is $\zeta=1.53 (1.51, 1.55)$ meaning that having received one e-mail from $j$ multiplies the odds that $i$ selects $j$ by 2.9, and ten e-mails by 39. Figure~\ref{fig:positions} shows the positions. The sender position is positive for 107 of the 120 employees, so that most senders rank the receivers in the same order and differ mainly in how strongly they discriminate. 
Most employees, among them 19 of the 20 most active, form a compact group above the centroid of the receiver positions, while the others spread out below it, with wide posterior regions. With one dimension the latent structure is therefore largely one of popularity. The
nine employees of the third quadrant are receivers whom the others do not select and who, as senders, choose receivers like themselves.

\subsection{Activity and position}
\label{sec:app-link}

The link reduces the residual standard deviation of the activity effects from 1.25 in $\calM_1$ to 0.84 and explains $R^2_a=0.57$ of their variation between employees, with a log Bayes factor of 33.6 against the model without link (Table~\ref{tab:results}). The evidence comes almost entirely from popularity: adding popularity to a link in excess self-affinity raises the log marginal likelihood by 39.3, whereas adding excess self-affinity to
popularity raises it by 2.7.

\begin{figure}[!htbp]
\centering
\appfig{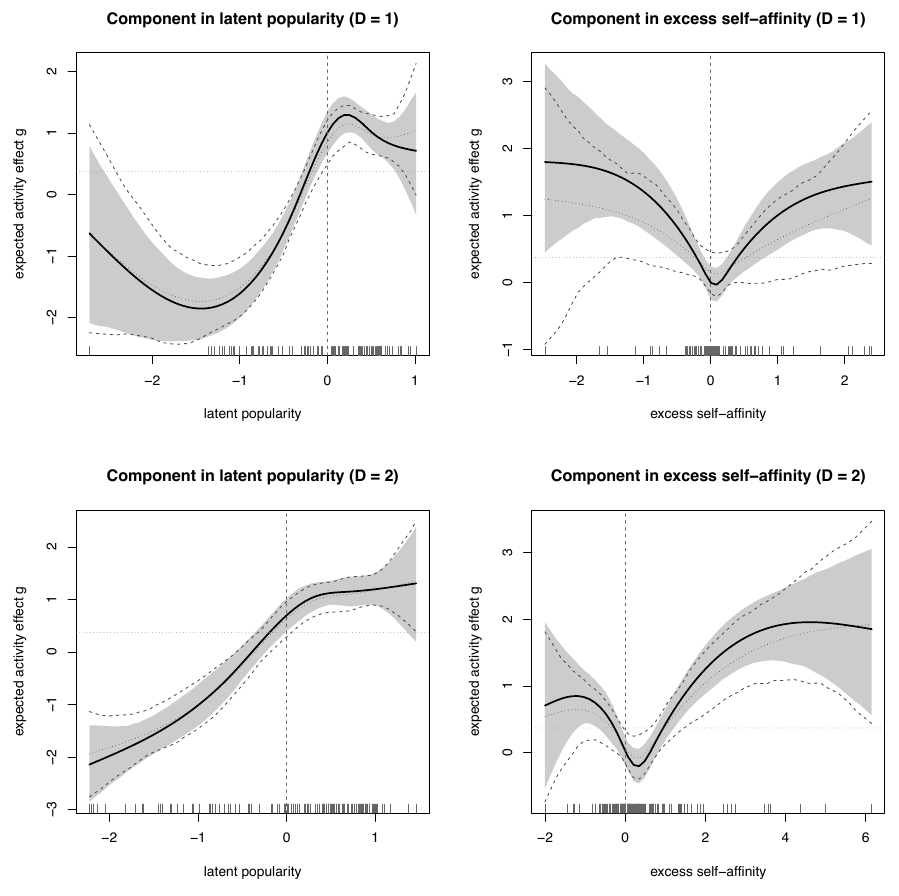}{width=\textwidth}
\caption{Components of the link in latent popularity (left) and excess self-affinity (right), with $D=1$ (top) and $D=2$ (bottom), with 95\% bands under the plug-in posterior (shaded) and the cut posterior (dashed).}
\label{fig:components}
\end{figure}

Figure~\ref{fig:components} shows the activity effects estimated without the link against the two summaries, and the two components of the link. The popularity component rises steeply, from about $-1.8$ for the least popular employees to about 1.3 for employees slightly above the average receiver, and does not rise further. Its contrast is $2.54 (2.10, 3.01)$ meaning that, other things being equal, the typical rate of an employee at the 90\% quantile of popularity is
about 13 times that of an employee at the 10\% quantile. The excess self-affinity component is lowest near zero and increases towards both extremes. However, this pattern is driven by a few employees, is roughly halved under the cut posterior, and is sensitive to the number of basis functions (Appendix~\ref{app:robust}).

The employees who write most are therefore those whom the others select as recipients, beyond the reciprocation of e-mails received (which $\zeta$ accounts for). This is the covariance between sender and receiver effects of the social relations model, for which a DyNAM with separate sub-processes has no parameter. The decomposition of self-affinity matters for this conclusion. A link in self-affinity alone is also supported, but much less (log Bayes
factor 21.6), because for most employees self-affinity is popularity rescaled by the strength of their preferences, which is not itself related to activity (Appendix~\ref{app:robust}). By itself, the link in self-affinity would suggest that employees who interact with their own kind are more active; the
decomposition shows that what matters is how strongly the others select them.

\begin{table}[t]
\centering
\caption{Main results with one, two and three latent dimensions. Choice
score: gain in the held-out choice score over the model with one dimension
fewer ($D=0$ is $\calM_0$), with standard error. Log Bayes factors: log
marginal likelihood of the rate module with each link relative to the model
without link, with $\sigma_a$ set in each model; `$\rho$ given $\pi$' and
`$\pi$ given $\rho$': gain from adding that component to a link in the other.
$R^2_a$: proportion of the variation of the activity effects explained by the
additive link. Contrast: level of the component at the 90\% minus level at
the 10\% quantile of its summary, with 95\% credible interval.}
\label{tab:results}
\smallskip
{
\begin{tabular}{@{}lccc@{}}
\hline
 & $D=1$ & $D=2$ & $D=3$\\
\hline
Choice score (gain over $D-1$) & 301 (29) & 175 (25) & 126 (25)\\[5pt]
Log Bayes factor & & & \\
\quad additive link ($\pi+\rho$) against no link & 33.6 & 39.5 & 37.1\\
\quad link in $\pi$ against no link & 30.9 & 28.5 & 31.0\\
\quad link in $r$ against no link & 21.6 & 14.4 & 13.6\\
\quad $\rho$ given $\pi$ & 2.7 & 11.0 & 6.1\\
\quad $\pi$ given $\rho$ & 39.3 & 45.0 & 43.9\\[5pt]
$R^2_a$ & 0.57 & 0.62 & 0.60\\
Contrast, $\pi$ & 2.54 (2.10, 3.01) & 2.71 (2.28, 3.16) & 2.97 (2.51, 3.45)\\
Contrast, $\rho$ & 0.15 ($-0.36$, 0.65) & 0.49 (0.03, 0.95) & 0.98 (0.50, 1.47)\\
\hline
\end{tabular}}
\end{table}

\subsection{Two and three dimensions}
\label{sec:app-D2}

With two dimensions the senders can favour different receivers, so that
excess self-affinity can vary across senders in more than sign and strength.
Table~\ref{tab:results} summarises the results with up to three dimensions,
fitted in the same way. The association with popularity hardly changes:
latent popularity with $D=2$ has correlation 0.90 with that for $D=1$, its
component has the same shape, and removing it from the link costs between 39
and 45 in log marginal likelihood. Excess self-affinity matters more: adding
it to popularity raises the log marginal likelihood by 11.0 with $D=2$,
against 2.7 with $D=1$, and its contrast is 0.49 $(0.03, 0.95)$. Its component
is again lowest near the median and rises on both sides, more steeply for
positive values, and unlike with one dimension this shape remains under the
cut posterior (Appendix~\ref{app:robust}). Given their popularity, employees who favour colleagues of their own kind more than the average sender also tend to be more active. This part of the association can only be
expressed with latent positions, but it is much smaller than the first, and
its size depends on the posterior mode of the choice module that is used.

\subsection{Robustness}
\label{sec:app-robust}

We summarise the robustness checks reported in Appendix~\ref{app:robust}.
The association with popularity is robust. Its contrast stays between 2.05
and 2.78 when the link is refitted with linear components, three or six basis
functions
it is $2.51\; (1.98, 2.97)$ under the cut posterior; it changes by at most 0.26 when the five employees with the most extreme summaries are removed from the link; and it is practically the same at the two best posterior modes of the choice module. The evidence for excess self-affinity with one dimension, in contrast, is weak: it changes sign with the number of basis functions and rests on a few
employees.

A concern specific to a link built on latent quantities is that the sender
position of an employee who sends few e-mails is shrunk towards the origin, which shrinks its excess self-affinity towards minus its popularity and could create an association with activity where there is none. We therefore applied the whole procedure to data simulated from the fitted model (Appendix~\ref{app:robust}). When the activity effects were reassigned at
random to the employees, which removes any association, the link explained at
most 8\% of the variation of the activity effects, against 57\% in the e-mail
data, and its log Bayes factor was always negative. When activity was related
to popularity only, which keeps the shrinkage, the log Bayes factor for excess
self-affinity given popularity was always negative, with one and with two
dimensions, whereas it is positive in the e-mail data. The  simulation study (Section~\ref{sec:simulation}) confirms that the procedure
recovers both components when they are present and does not find excess
self-affinity when only popularity is relevant.

\subsection{Prediction}
\label{sec:app-prediction}

We fitted the models to the first 80\% of the events and scored them on the
remaining 3,251. The latent positions improve the prediction of the
receivers: the choice score of $\calM_1$ exceeds that of $\calM_0$ by 301
(standard error 29), and further dimensions improve it by 175, 126 and 75.
The link does not improve the prediction of the sending times: the rate score
of $\calM_2$ is lower than that of $\calM_1$ by 2.1 (standard error 3.3), and
the difference is not distinguishable from zero in any quarter of the
employees ordered by activity. This is what \eqref{eq:shrink} leads one to
expect, since the rate of an employee who has sent many e-mails is determined
by that employee's own record. What the link contributes here is a
description of how activity varies with position, not a better forecast of
when employees send e-mails.

\section{Discussion}
\label{sec:discussion}

We have proposed a way to connect the two decisions of a DyNAM, when actors
act and whom they choose, through latent positions. Three elements make the
connection meaningful. First, the link is built on summaries of the positions
that the choice model identifies: latent positions are determined only up to
a group of transformations larger than the rotations, and summaries that
depend on the representation, such as the lengths of the vectors, partly
reflect the prior. Second, self-affinity, the only invariant summary of an
actor's own positions, splits into latent popularity and excess
self-affinity, which answer different substantive questions: how the others
see an actor, and how its preferences differ from theirs. Third, modular probabilistic 
inference keeps the positions a description of the choice of receivers, so
that the link relates two separately estimated characteristics of the same
actors.

In the e-mail data, the employees who write most are those whom the others
prefer as recipients. This association is strong and robust, and it is the
covariance between the sender and receiver effects which the link brings 
into DyNAMs. Preferences
that differ between senders add a smaller contribution once the latent space
has two dimensions: given their popularity, employees who favour their own
kind more than the average sender does tend to be more active. Without the
decomposition, a link in self-affinity would have attributed the association
to the employees' own preferences rather than to how the others see them. The
latent positions improve the prediction of the receivers, whereas the link
describes activity rather than forecasting it, because the activity of an
employee with a long record is determined by that record.

The analysis has limitations. The link describes an association and not an
effect of position on activity: employees who write a great deal may come to
occupy central positions. The positions, and so the link, describe what the
statistics in the model leave unexplained, and they would change with richer
statistics. The positions are constant within the observation window, the
dimension of the latent space is not selected. Finally, the public data contain no attributes of the employees,
so that we cannot say whether the latent structure reflects departments,
hierarchical levels or working groups; with such information the link could
be examined within and between units.

Natural extensions include positions that change over time, recorded
attributes in both sub-processes, and semi-modular inference, which lets a
controlled amount of information flow from the event times to the positions.
More generally, the link applies to any actor-oriented model in which the same
actors both choose and are chosen.

\section*{Data availability}

The e-mail data \citep{mic:pal:kaz11} are publicly available from the KONECT
network collection at \url{https://konect.cc/networks/radoslaw_email}. The R
package \textbf{ls.dynam}, which implements the methods is available at 
\url{https://github.com/acaimo/ls.dynam}.

\bibliographystyle{abbrvnat}

\newpage

\begin{appendices}

\section{Identification of the latent positions}
\label{app:ident}

Let $\mathcal A=\{1,\ldots,n\}$ be the set of actors, with sender positions
$\bfu_i\in\RR^D$ and receiver positions $\bfv_j\in\RR^D$. 
For any invertible $D\times D$ matrix $\bfA$ and any
$\bfw\in\RR^D$, consider the transformation $\bfu_i'=\bfA\bfu_i, \bfv_j'=\bfA^{-\top}(\bfv_j+\bfw)$ and let $\calG$ denote the group of all transformations of this form.

\begin{proposition}
\label{prop:ident}
Suppose that $n\ge 2D+2$, that any $D$ of the sender positions
$\bfu_1,\ldots,\bfu_n$ are linearly independent, and that any $D+1$
of the receiver positions $\bfv_1,\ldots,\bfv_n$ are affinely
independent. The latter condition means that, for any $D+1$ receiver
positions, the $D$ differences obtained by subtracting one of them
from the others are linearly independent.
Then
$\bfu_i'^\top(\bfv_j'-\bfv_k')
= \bfu_i^\top(\bfv_j-\bfv_k)$
for every $i$ and all $j,k\neq i$ if and only if
$(\bfU',\bfV')=g\cdot(\bfU,\bfV)$ for some $g\in\calG$.
\end{proposition}

\begin{proof}
\textit{``If'' part.}
Suppose that $(\bfU',\bfV')$ is obtained from $(\bfU,\bfV)$
through a transformation in $\calG$. Then
$$
\bfu_i'^\top(\bfv_j'-\bfv_k')
=(\bfA\bfu_i)^\top\bfA^{-\top}
  \bigl[(\bfv_j+\bfw)-(\bfv_k+\bfw)\bigr]
=\bfu_i^\top(\bfv_j-\bfv_k).$$
Note that $\bfu_i'^\top\bfv_j'=\bfu_i^\top\bfv_j+\bfu_i^\top\bfw$, so that
$\bfw$ adds to the linear predictor of sender $i$ a constant that does not
depend on the receiver $j$. Such a constant cancels in the choice
probabilities, which is why only the differences
$\bfu_i^\top(\bfv_j-\bfv_k)$ are identified.

\noindent
\textit{``Only if'' part.}
Now suppose that
$
\bfu_i'^\top(\bfv_j'-\bfv_k')
=
\bfu_i^\top(\bfv_j-\bfv_k)
$
for every $i$ and all $j,k\neq i$. We will show that there exist
an invertible matrix $\bfA$ and a vector $\bfw$ such that
$
\bfu_i'=\bfA\bfu_i,
\bfv_j'=\bfA^{-\top}(\bfv_j+\bfw).$

We choose any subset $J\subset\mathcal A$ containing $D+1$ actors,
and fix one actor $j_0\in J$. Define the $D\times D$ matrix
$\bfDelta_J
=
\bigl[\bfv_j-\bfv_{j_0}\bigr]_
{j\in J\setminus\{j_0\}},$
where the vectors inside the brackets are the columns of the matrix.
Define $\bfDelta_J'$ in the same way, using the primed receiver
positions.
By the assumption of affine independence, $\bfDelta_J$ is invertible.
For every actor $i\notin J$, the assumed equality applies to the
pairs $(j,j_0)$ for all $j\in J\setminus\{j_0\}$. Collecting these
$D$ equalities gives
\[
\bfDelta_J'^\top\bfu_i'
=
\bfDelta_J^\top\bfu_i.
\tag{1}
\]
There are $n-D-1\ge D$ actors outside $J$. By assumption, any
$D$ of their sender positions are linearly independent, so these
positions span $\RR^D$. Since $\bfDelta_J$ is invertible, the vectors
$\bfDelta_J^\top\bfu_i$ also span $\RR^D$.
Equation (1) shows that all these vectors belong to the image of
$\bfDelta_J'^\top$. Therefore, this image must be the whole of
$\RR^D$, which implies that $\bfDelta_J'$ is also invertible.
We can now solve Equation (1) for $\bfu_i'$:
\[
\bfu_i'=\bfA_J\bfu_i,
\qquad
\bfA_J=\bfDelta_J'^{-\top}\bfDelta_J^\top,
\qquad i\notin J.
\tag{2}
\]
Thus, for every subset $J$ of size $D+1$, we have found an
invertible matrix $\bfA_J$ that transforms the sender positions
of all actors outside $J$.

Now suppose that $J'$ is obtained from $J$ by replacing one actor
with another actor outside $J$. Then $J\cup J'$ contains $D+2$
actors. Since $n\ge 2D+2$, at least $D$ actors lie outside
$J\cup J'$.
For every such actor $i$, Equation (2) holds for both subsets,
and hence $\bfA_J\bfu_i=\bfu_i'=\bfA_{J'}\bfu_i.$
Choose $D$ of these actors. Their sender positions are linearly
independent by assumption. Since the two matrices agree on a basis of $\RR^D$, they must be equal:
$\bfA_J=\bfA_{J'}.$
Any two subsets of $\mathcal A$ containing $D+1$ actors can be
connected by a sequence of replacements of this kind. Consequently, all the matrices $\bfA_J$ are equal to a single invertible matrix $\bfA$.
For any actor $i$, we can choose a subset $J$ of size $D+1$
that does not contain $i$. Equation (2) then gives
\[
\bfu_i'=\bfA\bfu_i
\qquad\text{for every }i.
\tag{3}
\]

We now fix two distinct actors $j$ and $k$. For every actor
$i\notin\{j,k\}$, the assumed equality and Equation (3) give
\[
\bfu_i^\top
\left\{
(\bfv_j-\bfv_k)
-\bfA^\top(\bfv_j'-\bfv_k')
\right\}=0.
\tag{4}
\]
There are $n-2\ge D$ actors outside $\{j,k\}$. Their sender positions
span $\RR^D$, because any $D$ sender positions are linearly
independent. Therefore, the only vector orthogonal to all these
positions is the zero vector. Equation (4) consequently implies
that
\[
\bfv_j-\bfv_k
=
\bfA^\top(\bfv_j'-\bfv_k'),
\quad\textrm{or}\quad
\bfv_j'-\bfv_k'
=
\bfA^{-\top}(\bfv_j-\bfv_k).
\tag{5}
\]

It remains to show that receiver positions have the required form with a single vector $\bfw$. We fix actor $1$ and define
$\bfw=\bfA^\top\bfv_1'-\bfv_1.$
Taking $k=1$ in Equation (5), for every $j\neq 1$, gives
$\bfA^\top(\bfv_j'-\bfv_1')
=\bfv_j-\bfv_1.$
Rearranging and using the definition of $\bfw$, we obtain
$
\bfA^\top\bfv_j'=\bfv_j+\bfw.
$
The same equality holds for $j=1$ by definition. Hence
$
\bfv_j'=\bfA^{-\top}(\bfv_j+\bfw),$ for every $j.$ Together with Equation (3), this proves that
$(\bfU',\bfV')=g\cdot(\bfU,\bfV)$ for some transformation $g\in\calG$.
\end{proof}

The argument adapts that of \citet[Theorem~5.1]{and:rub56} for factor
analysis. Since self-directed events are excluded, $\bfu_i^\top\bfv_i$ never
enters the likelihood. The posterior is invariant to $g$ only if the prior is,
that is, only for rotations and reflections with $\bfw=\bfzero$, so that at a
posterior mode $\sum_j\hat\bfv_j=\bfzero$ and
$\sigma_u^{-2}\hat\bfU^\top\hat\bfU=\sigma_v^{-2}\hat\bfV^\top\hat\bfV$: the
prior fixes the location, scale and deformation of the configuration. Invariant
quantities depend on $(\sigma_u,\sigma_v)$ only through $\sigma_u\sigma_v$.

Self-affinity is the only invariant summary of an actor's own positions:
given the centroid, $r_i$ is a maximal invariant of $(\bfu_i,\bfd_i)$.
Invariant summaries that are not functions of $r_i$, such as $\pi_i$ and
$\rho_i$, depend on the positions of the other actors.

\section{Robustness checks for the e-mail data}
\label{app:robust}

\subsection{Bayes factors} The log Bayes factors of the main text compare
marginal likelihoods with $\sigma_a$ set separately in each model. Given the
value of $\sigma_a$ in the model with the link, the log Bayes factor for the
link with $D=1$ is 50.8 by the Savage-Dickey density ratio
\citep{ver:was95}, computed from the posterior draws, and 54.3 by the Laplace
approximation, and it is between 43.6 and 57.6 for prior standard deviations
of the link coefficients between 0.1 and 2. Computed in the same way, the log
Bayes factor for popularity given excess self-affinity is 55.0 and 59.0, and
that for excess self-affinity given popularity is 2.6 and 3.4, between 1.0
and 5.1 over the same prior scales; with $D=2$ the latter is 13.4 and 13.8,
between 5.9 and 15.2.

\subsection{Specification of the link} Table~\ref{tab:sensitivity} refits the
additive link with variants of its specification. The log Bayes factor for the
link is at least 35.3 and the contrast of the popularity component between
2.05 and 2.78, whereas the evidence for excess self-affinity given popularity
is positive with four basis functions per component but negative with three,
six or linear components. Removing from the link the five employees with the
most extreme values of each summary changes the popularity component by at
most 0.26, but the component in excess self-affinity by as much as 1.35. A
link in the selectivity $\phi_i=|u_i|\,s_v$, where $s_v$ is the standard
deviation of the receiver positions, which measures the strength of an
employee's preferences and is invariant, has a log marginal likelihood of
$-0.7$ relative to the model without link.

\begin{table}[h]
\centering
\caption{The additive link refitted with variants of the specification, $D=1$
(Laplace approximation to the rate module, $\sigma_a$ by maximum marginal
likelihood). Log Bayes factors, given $\sigma_a$, for the link against the
model without link and for excess self-affinity given popularity
($\rho\mid\pi$). Contrast: level at the 90\% minus level at the 10\%
quantile of each summary.}
\label{tab:sensitivity}
\smallskip
{
\begin{tabular}{@{}lrrrrrrr@{}}
\hline
 & & & \multicolumn{2}{c}{$\log\mathrm{BF}$} & & \multicolumn{2}{c}{Contrast}\\
Specification & $K$ & $\sigma_a$ & link & $\rho\mid\pi$ & $R^2_a$ & $\pi$ & $\rho$\\
\hline
Splines, $K_l=4$ (main) & 8 & 0.84 & 54.3 & 3.4 & 0.57 & 2.51 & 0.14\\
Linear components & 2 & 0.96 & 35.3 & $-2.3$ & 0.41 & 2.05 & $-0.06$\\
Splines, $K_l=3$ & 6 & 0.89 & 42.9 & $-1.0$ & 0.50 & 2.44 & $-0.03$\\
Splines, $K_l=6$ & 12 & 0.83 & 49.1 & $-0.1$ & 0.58 & 2.76 & 0.01\\
\hline
\end{tabular}}
\end{table}

\subsection{Cut posterior} With $D=1$ the contrast of the popularity component
is $2.51\; (1.98, 2.97)$ under the cut posterior, against $2.54\; (2.10, 3.01)$
under the plug-in posterior. The level of the component in excess
self-affinity at the 5\% and 95\% quantiles of $\rho_i$ exceeds that at the
median by $1.17\; (0.51, 1.83)$ and $1.19\; (0.52, 1.83)$ under the plug-in
posterior, but by $0.63\; (-0.08, 1.47)$ and $0.63\; (-0.31, 1.50)$ under the
cut posterior. With $D=2$ the level at the 90\% and 95\% quantiles exceeds that at the median by $1.32\; (0.85, 1.79)$ and
$1.84\; (1.27, 2.42)$ under the plug-in posterior, and by $1.02\; (0.46, 1.59)$ and $1.47\; (0.80, 2.21)$ under the cut posterior.

\subsection{Posterior modes of the choice module} With $D=1$ the optimisation
from 31 starting configurations, 30 random and one spectral, found eight
distinct modes (Table~\ref{tab:modes}). The best was
reached 4 times and the second-best, with a log-posterior 6.8 lower, 11
times; the two describe similar structures and give the same conclusions
about the link, whereas the other modes fit the choices far less well, and
the sampler stays near the best mode. With $D=2$ the best mode was reached by
11 of 31 starts; at the ten other modes the log Bayes factor for excess
self-affinity given popularity is between $-5.0$ and 14.0, whereas the
contrast of the popularity component stays between 2.42 and 2.77 at all of
them except the worst. With $D=3$ the interval for the contrast of excess
self-affinity excludes zero at five of the six modes found.

\begin{table}[ht]
\centering
\caption{Posterior modes of the choice module reached from 31 starting configurations,
$D=1$, and the link fitted at each of them (Laplace approximation, summaries
evaluated at the mode). Starts: number of starting configurations that
reached the mode ($^\ast$ including the spectral start). $\Delta$:
log-posterior of the best mode minus that of the mode. Correlations with the 
best posterior mode of $\pi_i$, of $\rho_i$ and of the matrices $\bfU\bfD^\top$ of 
centred latent affinities $\bfu_i^\top\bfd_j$, where $\bfD$ has rows $\bfd_j^\top$.}
\label{tab:modes}
\smallskip
{
\begin{tabular}{@{}clrrrrrrrrr@{}}
\hline
 & & & \multicolumn{3}{c}{Correlation} & \multicolumn{2}{c}{$\log\mathrm{BF}$} & & \multicolumn{2}{c}{Contrast}\\
Mode & Starts & $\Delta$ & $\pi$ & $\rho$ & $\bfU\bfD^\top$ & link & $\rho\mid\pi$ &
$R^2_a$ & $\pi$ & $\rho$\\
\hline
1 & 4 & 0.0 & 1.00 & 1.00 & 1.00 & 54.6 & 4.1 & 0.57 & 2.50 & 0.10\\
2 & 11$^\ast$ & 6.8 & 0.87 & 0.82 & 0.78 & 52.5 & 4.0 & 0.56 & 2.41 & 0.27\\
3 & 3 & 116.8 & 0.34 & 0.41 & 0.18 & 28.9 & $-5.4$ & 0.43 & 2.12 & 0.35\\
4 & 8 & 150.6 & 0.48 & 0.36 & 0.31 & 26.8 & $-4.4$ & 0.42 & 2.12 & 0.32\\
5 & 2 & 219.6 & 0.45 & 0.04 & 0.25 & $-2.9$ & $-6.1$ & 0.12 & 0.92 & 0.11\\
6 & 1 & 252.4 & 0.30 & 0.31 & 0.07 & $-2.3$ & $-3.2$ & 0.14 & 0.60 & $-1.03$\\
7 & 1 & 253.5 & 0.25 & 0.26 & 0.05 & $-4.5$ & $-4.4$ & 0.11 & 0.54 & $-0.87$\\
8 & 1 & 347.2 & 0.12 & 0.32 & 0.02 & 2.1 & $-6.6$ & 0.20 & 0.32 & $-0.25$\\
\hline
\end{tabular}}
\end{table}

\subsection{Calibration by simulation} We simulated 20 replicate data sets
from the fitted model with $D=1$, with the estimated positions and
coefficients, and applied the whole procedure to each, computing the
summaries at the posterior mode. With the estimated activity effects randomly
reassigned to the employees, which removes any association while keeping the
positions and the amount of data per employee, the log Bayes factor for the
link was between $-13.5$ and $-7.2$, $R^2_a$ at most 0.08, and the contrasts
of the two components between $-0.39$ and 0.78 and between $-0.61$ and 0.52,
against 54.6, 0.57, 2.50 and 0.10 for the e-mail data computed in the same
way. With activity effects simulated from the fitted link in popularity
alone, with randomly reassigned residuals, the log Bayes factor for excess
self-affinity given popularity was between $-8.3$ and $-4.7$, against 4.1. The
same calibration with $D=2$ gave values between $-8.3$ and $-0.8$, against
16.6.

\end{appendices}

\end{document}